\documentclass[a4paper]{llncs}
\usepackage[utf8]{inputenc}
\usepackage{textcomp}

\usepackage{graphicx}
\usepackage{amsmath}
\usepackage[version=4]{mhchem}
\usepackage{siunitx}
\usepackage{longtable,tabularx}
\usepackage{blindtext}
\usepackage{mathrsfs}
\usepackage{caption}
\usepackage{adjustbox}
\usepackage{xargs}
\usepackage{relsize}
\usepackage{bm}
\usepackage[pdftex,dvipsnames]{xcolor}
\usepackage[colorinlistoftodos,prependcaption,textsize=tiny]{todonotes}
\usepackage{array}
\usepackage{booktabs}

\usepackage{url}
\usepackage{breakurl}
\usepackage[breaklinks]{hyperref}
\usepackage{cleveref}
\usepackage[T1]{fontenc}

\makeatletter
\newcommand{\@chapapp}{\relax}%
\makeatother
\usepackage[page, title,header]{appendix}
\usepackage{tikz}
\usetikzlibrary{fit, positioning}

\newcommand{\T}[1]{\ensuremath{\mathcal{T}_{#1}}}

\newcommand\period{\ensuremath{\mathcal{B}}}
\newcommand\phoenix{\ensuremath{\mathcal{P}}}

\newcommand\delay{\ensuremath{\mathcal{D}}}
\newcommand\allrequests{\ensuremath{\mathcal{R}}}
\newcommand\request{\ensuremath{r}}
\newcommand\funds{\ensuremath{\mathcal{F}}}
\newcommand\unlock{\ensuremath{\mathcal{U}}}
\newcommand\lastid{\ensuremath{\mathcal{L}}}

\newcommandx{\unsure}[2][1=]{\todo[linecolor=red,backgroundcolor=red!25,bordercolor=red,#1]{#2}}
\newcommandx{\change}[2][1=]{\todo[linecolor=blue,backgroundcolor=blue!25,bordercolor=blue,#1]{#2}}
\newcommandx{\info}[2][1=]{\todo[linecolor=OliveGreen,backgroundcolor=OliveGreen!25,bordercolor=OliveGreen,#1]{#2}}
\newcommandx{\improvement}[2][1=]{\todo[linecolor=Plum,backgroundcolor=Plum!25,bordercolor=Plum,#1]{#2}}
\newcommandx{\thiswillnotshow}[2][1=]{\todo[disable,#1]{#2}}

\newcommand{\bpar}[1]{\paragraph{\textup{\textbf{#1}}.} }

\title{Phoenix: A Formally Verified Regenerating Vault} 

\author{Uri Kirstein, Shelly Grossman, Michael Mirkin, James Wilcox, Ittay Eyal, Mooly Sagiv}

\institute{Technion, Israel Institute of Technology and Certora}

\begin{document}

\sloppy

\maketitle
\begin{abstract} 
An attacker that gains access to a cryptocurrency user's private keys can perform any operation in her stead. 
Due to the decentralized nature of most cryptocurrencies, no entity can revert those operations. 
This is a central challenge for decentralized systems, illustrated by numerous high-profile heists. 
\emph{Vault contracts} reduce this risk by introducing artificial delay on operations, allowing abortion by the contract owner during the delay. 
However, the theft of a key still renders the vault unusable and puts funds at risk. 

We introduce \emph{Phoenix}, a novel contract architecture that allows the user to restore its security properties after key loss. 
Phoenix takes advantage of users' ability to store keys in easily-available but less secure storage (\emph{tier-two}) as well as more secure storage that is harder to access (\emph{tier-one}). 
Unlike previous solutions, the user can restore Phoenix security after the theft of tier-two keys and does not lose funds despite losing keys in either tier. 
Phoenix also introduces a mechanism to reduce the damage an attacker can cause in case of a tier-one compromise.

We formally specify Phoenix's required behavior and provide a prototype implementation of Phoenix as an Ethereum contract.
Since such an implementation is highly sensitive and vulnerable to subtle bugs, we apply a formal verification tool to prove specific code properties and identify faults. 
We highlight a bug identified by the tool that could be exploited by an attacker to compromise Phoenix.
After fixing the bug, the tool proved the low-level executable code's correctness.

\end{abstract}


    \section{Introduction}

\emph{Cryptocurrency} protocols allow users to exchange virtual currency tokens. 
User funds are secured with cryptographic tools, and each user controls her funds by issuing cryptographically-signed orders called \emph{transactions}. 
The raison d'être of most cryptocurrencies is their \emph{decentralization}: no central party controls the state of their system and its progress~\cite{narayanan2016bitcoin,oakland_crypto_book}.
Cryptocurrencies are usually implemented on top of a blockchain~\cite{nakamoto2019bitcoin}, an immutable, tamper-free ledger of transactions.
That is, once a transaction has taken place, it cannot be reverted~\cite{bitcoin2020irreversible}, even for those currencies that are centrally controlled~\cite{androulaki2018hyperledger_fabric,libra_2019}. 
This is in stark difference to traditional banking~\cite{bank_transaction_reversion} and similar systems~\cite{credit_fraud_reversion}. 
There, if a mistaken transaction is detected due to fraud, theft, or a mistake, the system operator can revert it and undo the harm. 

The implication is that private key security is a prominent concern that accompanies the usage of cryptocurrencies. 
In numerous events, funds were lost due to mistakes such as accidental transfers to wrong destinations~\cite{wrong_address_bad,wrong_address_bitcoin,investopedia2020lost,cointelegraph2020lost}.
Moreover, private keys are highly coveted by hackers, as they allow for immediate theft of funds; indeed, theft cases on all major cryptocurrencies have been reported, worth millions of USD~\cite{tan_timeline:_2018,tsihitas_biggest_2019}.
In all such cases, users have no recourse. 

Various approaches have been suggested to improve user-side security. 
One is to require multiple keys per operation~\cite{multisig}; thus, only if several keys are compromised can the funds be stolen. 
However, if some of the keys are lost, the funds are also lost. 
\emph{cold storage}~\cite{antonopoulos2018mastering} is the practice of storing a key on a device not connected to the internet, possibly a piece of paper.
However, cold storage is more cumbersome to access, making it inappropriate for routine operations.

M{\"o}ser, Eyal, Sirer et al.~\cite{moser2016bitcoin} (MES) introduced the idea of a \emph{cryptocurrency vault}. 
A vault is a \emph{smart contract}~-- an automaton implemented by the underlying cryptocurrency that controls funds and can execute arbitrary logic. 
For funds in the vault, money withdrawal is delayed by an amount of time set by the user, e.g., 24 hours. 
During this period, the owner can cancel the transaction.
The transaction is not complete until the delay time has elapsed, but during this time, it allows the rightful owner to abort the transaction if it was issued by mistake or by an attacker after a theft.
In contemporary work, Bartoletti, Lande, Zunino et al.~\cite{bartoletti2020bitcoin} (BLZ) propose using a different key for transaction abortion, one that is harder to steal. 
However, in both works, once any key is lost or stolen, the vault contract is unusable.
We review previous and related work in \Cref{sec:related}. 

In this paper, we present \emph{Phoenix Vault}, a novel smart-contract whose security can be restored despite the theft of private keys, resurrected like the mythical Phoenix. 
Phoenix employs two \emph{tiers} of privileged keys, which can be added and removed under certain limitations.
It takes advantage of the fact that each user can use two different types of storage: accessible storage, which is vulnerable, and secure storage, which is hard to access.
\emph{Tier-two} addresses are stored in accessible storage; each is used to initiate withdrawals and abort a withdrawal it mistakenly initiated. 
\emph{Tier-one} addresses are stored in secure storage; they are used to abort any transaction (issued by an attacker), add tier-one and tier-two addresses dynamically, and remove tier-two addresses. 
In case of a theft of a tier-two address, the Phoenix can then be recovered by replacing the stolen key with a new one.
An attacker compromising a tier-one address can initiate withdrawals with several tier-two addresses they generate.
Phoenix introduces a novel mechanism to avoid attrition warfare in this scenario: the user can put the contract in a timed lock-down state, such that no withdrawals are possible. 
We overview Phoenix's specification and underlying model in \Cref{sec:ModelAndSpec}.

We implemented Phoenix on the Ethereum blockchain~\cite{buterin2014ethereum_whitepaper,ethereum_yellow_paper}. 
Ethereum's blockchain supports arbitrarily complex contracts~\cite{turing_completeness} written in a compiled JavaScript-like language called Solidity~\cite{solidity_0.6.6}. 
We implemented Phoenix in Solidity and executed it in a local testnet to prove its viability. 
Phoenix is implemented in two parts. 
A library implements the ledger holding all withdrawal requests, and the main contract implements the maintenance of the two tiers of addresses, the operations, and their authorization.
We discuss our implementation and code architecture in \Cref{Se:Architrcture}.

Phoenix's development was accompanied by reasoning about high-level properties describing its precise specification.
This reasoning requires mathematically formulating the exact high-level properties of the code and developing techniques for proving that the actual code adheres to the specification.
We use the Certora Prover~\cite{certora_homepage} to verify that the low-level EVM bytecode~\cite{ethereum_yellow_paper}, as compiled from Solidity, obeys the high-level specification.
We used the tool to automatically find subtle bugs and prove their absence.
For example, the tool uncovered a bug that could be exploited to perform a denial of service attack.
For details, see~\Cref{sub:verification}.

Our contributions can be summarized as follows: 
\begin{itemize}
\item A model that formally distinguishes accessible and secure storage.
\item A formal safety specification of a novel Phoenix vault.
\item A mechanism for retrieving funds even if all withdrawal keys were stolen or lost.
\item A prototype implementation of a Phoenix vault on the Ethereum network.
\item Formal reasoning about the bytecode implementation. 
\end{itemize}

Our Solidity code, test file, and specification files are published as open-source code at the link found in \Cref{Se:Architrcture}.

We conclude in \Cref{sec:conclusion}, where we also show potential applications of Phoenix for corporate access management and a will contract. 


    \section{Background and Related Work} \label{sec:related} 
    \subsection{Lost Keys Problem}

By and large, in cryptocurrencies, spending of funds is done via private keys~\cite{cryptocurrencies_use_private_keys}. 
Whoever knows the key has complete control over the funds and can instantly move the funds by spending them however they want.
The tamper-free nature of the blockchain makes this transaction final~\cite{wrong_address_bad,wrong_address_bitcoin}. 
In the traditional financial system, an erroneous transfer to the wrong recipient by a stolen credit card or wire fraud can be reverted by the credit company or the legal system to recover the funds~\cite{bank_transaction_reversion,credit_fraud_reversion}. 
However, no authority, regulatory body, or institutional apparatus can change the history of the blockchain~\cite{noauthor_immutability-io/vault-ethereum_2020}. 
Therefore, such money transactions are irrecoverable without an agreement to change the currency's protocol by a large portion of the currency's participants~\cite{moser2016bitcoin}.

In the history of Ethereum, only one precedent exists where a significant portion of the community agreed to revert a transaction. 
This incident was a reaction to \emph{The DAO}~\cite{antonopoulos2018mastering} smart contract that held 3.6 million ether, which almost everyone believed had been stolen. 
This move was controversial and spawned a \emph{fork} of Ethereum into two different networks - Ethereum and Ethereum Classic~\cite{antonopoulos2018mastering}. 
Retrieving stolen money is, therefore, impractical for an individual.

Due to the irreversibility of transactions, cryptocurrency private keys are attractive theft targets.
Once an adversary successfully transfers funds to their possession, nothing can be done to revert the action, even if the breach was noticed moments after the attack. 
There have been several such attacks on most major cryptocurrencies worth millions of USD.
It is noteworthy that not just individuals suffer from this problem, but large companies and major players fall prey to these attacks too~\cite{tan_timeline:_2018,tsihitas_biggest_2019}. 

Key theft is a part of the more general problem of the key being inaccessible. 
A password's storage device might be damaged, making key retrieval impossible.
If no storage device is used, there is a total reliance on the owner's memory, who might forget the password.
There is also the risk of the owner passing away.
If a wallet owner is deceased without sharing the key beforehand, the wallet's contents will remain forever inaccessible~\cite{rice2019deceased}.

A study from 2019 estimated 20\% of all mined Bitcoin has been lost for various reasons~\cite{krause_fifth_2018}. 
Consequentially, safely storing keys is among the most concerning security vulnerabilities from the user's perspective.
It limits a consumer's trust in cryptocurrency~\cite{devries2016analysis}. 
The lack of adequate private-key security solutions is likely a factor that has prevented widespread adoption of the technology~\cite{esmaeilzadeh2019individuals}.

    \subsection{Related Work}

The simplest solution to the problem is to use a cryptocurrency network with reversible transactions.
Gurcan et al. have proposed a cryptocurrency model that allows the cancellation of transactions~\cite{gurcan2018cancellation}.
They base their model on a bitcoin-like blockchain.
However, to our knowledge, no cryptocurrency network in use today allows the reversal of transactions.
Until such technology is developed, secure key management is a necessity.

Different approaches to key management offer different trade-offs between availability, security, and convenience~\cite{narayanan2016bitcoin}. 
The most convenient and available solution is storing the keys on a local device, be it a phone or a computer. 
However, the device is highly vulnerable to attacks; it can be stolen or infected by malware such as \emph{viruses}, \emph{worms}, \emph{Trojan horses} and \emph{ransomware}~\cite{nakamoto2018w}.

A common practice for securely storing cryptocurrency funds is by using \emph{cold storage}~\cite{antonopoulos2018mastering}. 
In cold storage, the key is stored on a device not connected to the internet. 
However, regular interaction with the key is still required~\cite{moser2016bitcoin}. 
While money can be deposited to the cold storage using public knowledge, spending must be done using the private key. 
These interactions are points of vulnerability, during which an attacker can get hold of the key.
Additionally, cold storage is only as secure as the storage device's physical security.
Protecting a physical device is not necessarily easier than protecting a digital asset.

An alternative to cold storage is never to use a physical device at all and remember the password. 
One method to do so is to use a \emph{Brain Wallet}, a way to create keys from an easy-to-remember passphrase. 
Several companies provide this service, such as Keybase's WarpWallet~\cite{krohn_keybase/warpwallet_2020,krohn_warpwallet_nodate}. 
Brain Wallets introduce the risk of the passphrase being forgotten, as it is not stored anywhere. 
Brain Wallets are also susceptible to \emph{dictionary attacks}. 
Attackers can perform offline guesses of word combinations from a word list to test them against a candidate password~\cite{brain_drain}. 
Many secret keys can be targeted at once, mitigating the costs to launch such attacks.
Perhaps due to their inherent risk, few brain wallets were used in practice. 
Vasek et al. estimated the number of Brain Wallet in active use in 2016 to be less than 1,000~\cite{brain_drain}.

Many companies have tried to solve the problem of secure key storage via various methods. 
Curv~\cite{noauthor_curv_nodate} replaces keys with multi-party computational protocols (MPCs). 
Another project named Ethereum Vault~\cite{noauthor_immutability-io/vault-ethereum_2020} implements a plugin for existing HashiCorp Vault software for generic data secure storage as a vault for Ether. 
The closest form of a commercial solution to the one we present in this paper is CoinBase Vault~\cite{noauthor_coinbase_nodate}, which also features time delayed withdrawals. 
The underlying problem with all of them is the need to trust a third party not to save private keys for malicious purposes or sell them. 
Even if the user does trust the company providing key security services, the user cannot verify their products' quality and security against adversaries' attempts. 
Our Vault is an open-source code that was formally verified.

\bpar{Vault contracts}
The basic \emph{vault contract} structure was first proposed by MES~\cite{moser2016bitcoin}. 
It set the foundation for having withdrawals delayed in time instead of happening immediately. 
In their idea, the vault's key can be used not just to withdraw money, but also to abort withdrawal requests that have not yet been waiting for the entire delay period.
This scheme ensures that as long as the key's owner monitors the vault once every period equal to the length of the delay, no unauthorized withdrawals can materialize. 
The user can prevent transfers to wrong addresses if the error is spotted before the delay has passed.
One of the solution's problems is that it relies on a single, frequently used key.

BLZ vaults~\cite{bartoletti2020bitcoin} are an improved design of the contract, introducing two different keys. 
The first key is used for withdrawals, and the second key is used for the cancellation of withdrawals. 
Cancellations are rare events that only happen when the user erroneously sent a withdrawal request, or if the first key was stolen. 
The second key is far less used and exposed than the first, increasing its security. 
However, this solution offers no means of recovery once a key has been compromised. 
A stolen key renders the vault useless - either the owner has to cancel withdrawals in perpetuity as an attrition war that costs both sides gas, or the owner cannot withdraw the money. 
Phoenix has a solution to completely recover from the loss or theft of the low privilege key. 
It features a mechanism to reach a stationary position that does not incur any further loss of money under any key theft. 
Unlike BLZ vaults, our Phoenix design was implemented on an existing blockchain network.

Swambo et al. have proposed a custody protocol based on a vault contract for Bitcoin to control funds from loss and theft~\cite{swambo2020custody}.
They implemented their vault contract using pre-signed transactions and secure key-deletions.
They are the only ones to our knowledge that have published a vault contract's code online and tested it.
However, their solution requires numerous hardware devices.
Phoenix requires no external devices and is economical for the storage of even modest amounts.
Additionally, their solution assumes there is no infiltration; the operators of the protocol are honest.
Our proposition does not assume the same and mitigates the risk even when an adversary has the same capabilities as Phoenix's owner.



    \section{Model and Specification Overview}\label{sec:ModelAndSpec}

In this section, we overview the model and specification of Phoenix. 
A formal specification is presented in \Cref{sub:FormalSpec}. 


        \subsection{Model} 

\bpar{Users and the blockchain} 
The system comprises users with access to a cryptocurrency blockchain. 
Each user can generate a private-public key pair, and the public key serves as an \emph{address}. 
The blockchain assigns an amount to each address. 
At any time a user can query the blockchain for the amount assigned to an address; this is the \emph{cryptocurrency} controlled by that address. 
A user can issue an order to transact cryptocurrency from one address to another. 
Such an order is called a \emph{transaction}. 
It is valid only if signed with the sending address's private key, assuring only the cryptocurrency owner can send it. 

The blockchain aggregates valid transactions from the users and processes them in groups called \emph{blocks} (hence, blockchain) at set intervals. 

Users can also issue transactions that initiate and interact with so-called \emph{smart contracts}. 
These are logical stateful automatons that also control cryptocurrency~-- they can receive it and send it. 

        \bpar{Storage and adversary} 
Each user can use two types of storage for her private keys. 
The first is \emph{secure storage}, which is highly secure, but also cumbersome to access. 
This could be a physical bank safe. 
The second is \emph{accessible storage}, which can be conveniently accessed, but is more prone to theft. 
This could be a mobile phone application, for example. 

An adversary might gain access to either storage type, though the former is much less likely. 
Moreover, she might do that covertly, without the victim realizing her information was accessed. 

We consider two attack types. 
In an attack of \emph{Type I}, the adversary obtains one or more addresses from the secure storage and tries to transact funds from the control of a user to herself. 
In an attack of \emph{Type II}, the adversary obtains one or more addresses from the accessible storage and tries the same.
In practice, this attack is more common as accessible storage is (by definition) more vulnerable. 

    \bpar{Goal}
Our goal is to design a contract that enables a user to fully use it with accessible-storage access only and secure the user's funds in the face of an attack. 
In case of a Type~I attack, the attacker should not be able to steal any funds, but she may prevent the user from moving her funds. 
However, in case of a Type~II attack, the victim should be able to revoke the stolen keys. 
Moreover, in case of benign loss of accessible keys, the user should be able to revoke the lost keys and replenish the contract with new ones.
Only if all privileged keys of all types are lost should the contract's funds be unreachable.


        \subsection{Specification Overview} 

Based on the goal stated above, we can now present the specification of a Phoenix vault contract. 
The formal details are in \Cref{sub:FormalSpec}. 

Any user on the network can send cryptocurrency to the Phoenix contract, thus increasing its funds. 
Phoenix contains two sets of addresses: privileged \emph{tier-one} addresses and less-privileged \emph{tier-two} addresses.
Users interact with Phoenix by performing \emph{actions}, namely transactions that interact with the contract. 
Most actions are restricted only to addresses belonging to a particular tier, privileged to perform that action. 
Users are to store tier-one addresses in secure storage and tier-two addresses in accessible storage. 
Roughly, tier-two addresses are used for withdrawal, and tier-one addresses are used to abort malicious withdrawals and update the address sets. 

            \bpar{Withdrawal}
Tier-two addresses are used to withdraw money from Phoenix to an arbitrary address.
As money withdrawal is an everyday action, users can keep their tier-two private keys in accessible storage.
Given a withdrawal transaction, Phoenix withholds the funds for a predetermined delay duration before it can be transferred. 
During the delay duration, only a tier-one address or the same tier-two address that initiated the transfer can cancel the withdrawal and prevent the transfer.
Since cancellation is only performed due to attacks or benign errors, they are infrequent, and the overhead of accessing tier-one addresses does not affect regular operation.


            \bpar{Addresses}
We use the same primary key functions as BLZ vaults~\cite{bartoletti2020bitcoin}.
We add two innovations.
The first is allowing the existence of several keys of each tier.
The second is having dynamic keys: the user can add addresses to both tiers and remove tier-two addresses at will. 
This allows using Phoenix for various use-cases, such as serving a company with multiple workers or function as a smart will.
We expand on these use cases in \Cref{sec:conclusion}.
Besides expanded usability, these two innovations play a vital role in the wallet's ability to recover from attacks, as will be demonstrated next.

We want to minimize the usage of tier-one keys.
Therefore, we forbid an address to have both tier-one and tier-two privileges.
A tier-two address can cancel its own withdrawals, in cases of an erroneous withdrawal.

A single tier-one address must be assigned at Phoenix's creation time.
A tier-one address can add other tier-one and tier-two addresses.
A tier-one address cannot be removed from Phoenix's tier-one addresses set; else, an attacker could cause the creator of Phoenix to lose control over it.
\Cref{App:Tiers} summarizes the differences between the two privileges.


            \bpar{Key Loss Recovery}
Assuming no malicious parties, the user can also recover the funds in Phoenix if a key became inaccessible or was forgotten.
If a tier-two key was lost, but the user still possesses a tier-one key, they can generate a new network address and add tier-two privileges to it.
If a tier-one address is forgotten or lost, but access to a tier-two key is maintained, the owner can still withdraw funds from Phoenix, perhaps to a new Phoenix they have created. 
Only if all tier-one and tier-two addresses are lost or forgotten will Phoenix's funds be inaccessible.

            \bpar{Type-I Attack Recovery Mechanism}
In an attack of \emph{type I}, the adversary obtains a tier-one address.
Using it, the adversary can add tier-two privileges to network addresses they generated and use those tier-two addresses to request withdrawals from Phoenix.
Even under this attack, the adversary does not gain control of the funds secured by Phoenix.
As long as the user also has access to their tier-one address, they can cancel all the adversary's withdrawals.
However, the adversary can do the same to all legitimate withdrawals, rendering Phoenix useless from the user's point of view.
Nevertheless, the adversary does not directly profit from this attack, which by itself should deter thieves from trying to steal the Phoenix's user addresses.

A possible incentive for a type I attack could be to initiate an \emph{attrition warfare}.
The adversary can make frequent withdrawals, forcing the rightful user to cancel them before the delay is over, incurring time and money. 
The user must be eternally vigilant as each request that evades abortion may empty the funds stored in Phoenix.

To alleviate this problem and reach a stable state, Phoenix features a time lock mechanism: a tier-one address can lock Phoenix and set a time in the future in which it will be unlocked.
While Phoenix is locked, all transfers will fail.
A lock's opening time can only be postponed.
The resulting equilibrium is a lock with a timestamp far in the future, denying both parties the ability to withdraw money but requiring no interaction from either party to maintain the status quo.

            \bpar{Type-II Attack Recovery Mechanism}
If an adversary possesses a tier-two address, she can do two things: transfer funds to her own address and abort withdrawals issued by the user with that tier-two address. 
A tier-one address can remove that tier-two address from Phoenix's tier-two addresses set, then remove all the malicious withdrawals it initiated one by one. 
Rather than making this a two-step process, we specify that removing a tier-two address also cancels all of its previous withdrawals waiting for the delay to pass. 
This significantly reduces the number of accesses necessary to the tier-one key.
The user can then use the tier-one address to register a new tier-two address.
This mechanism provides full recovery in case of a type-II attack, restoring the security to its state before the theft took place. 

A tier-one address can cancel all pending withdrawals with one action. 
This ability is useful if multiple tier-two addresses were stolen, or if a tier-one was compromised and created several malicious tier-two addresses.
It reduces the number of actions required by the secure tier-one key to one.

            \bpar{Delay}
Phoenix's delay value is set at creation time and cannot be changed.
As noted by Swambo et al.~\cite{swambo2020custody}, the choice of the delay's length allows the user to choose their desired balance between accessibility and security. 
A short delay means less time for the user to respond to a breach and increases the risk of an unauthorized transaction.
On the other hand, a long delay means that the sender and the recipient have to wait longer for every payment, reducing accessibility.


    \section{Architecture and Implementation}
    \label{Se:Architrcture}

We implemented Phoenix in the Solidity language~\cite{solidity_0.6.6}, which compiles to a smart contract~\cite{buterin2014ethereum_whitepaper} that runs on the Ethereum Virtual Machine (EVM)~\cite{ethereum_yellow_paper}. 

        \bpar{Initialization}
The contract initiates two sets, for tier-one and tier-two addresses. 
The user that creates the contract initially assigns a single tier-one and a single tier-two address. 
By default, the tier-two address is the address that sent Phoenix's constructor transaction, for convenience.
The tier-one address at contract creation must be different from the address that has sent Phoenix's constructor transaction, to minimize its exposure. 
The user also chooses a delay interval~\delay\ for withdrawals.  

        \bpar{Withdrawal}
The EVM does not directly support the delayed execution of operations. 
Each transaction runs atomically at a single point in time. 
Therefore, withdrawal is performed in two steps. 
First, a transaction made by a tier-two address initiates a transfer \emph{request} providing a destination address. 
Later, another transaction, a \emph{withdrawal claim}, completes the operation and retrieves the funds.
Any key, even unprivileged, can issue withdrawal claims; in fact, it is better not to use a privileged key for this purpose to reduce exposure.
The contract makes sure that~$D$ blocks have passed since the matching request of the claim.
Block count is a good proxy for time, as blocks are generated at exponential intervals with a mean of~15 seconds~\cite{ethereum_yellow_paper}. 

        \bpar{Code architecture}


The byte code of the contract was too large to be deployed in one Ethereum transaction, exceeding the maximal gas limit~\cite{reynaldo_ethereum_2020} (a measure of contract overhead) of a block on the Ethereum network. 
It, therefore, comprises a Ledger library and the main contract. 
As a library, the Ledger must be deployed before the main contract, and the main contract links to its address. 
Libraries~\cite{solidity_contracts_documentation} cannot have state variables, cannot receive Ether, and cannot be destroyed.
The Ledger must be deployed once, and then every Phoenix contract can link to its address. 
The Ledger is responsible for recording withdrawal requests. The main contract holds the balance and is mainly responsible for managing access restrictions according to the user tiers.

Phoenix maintains a map (hash table) from user addresses to their privileges.
Privilege examination is a frequent action because most Phoenix interactions are limited to a particular privilege only. 
Therefore, ensuring it would be done in an average runtime of $O(1)$ is a high priority.
Using a map allows adding and deleting addresses in $O(1)$ average time, and guarantees that an address isn't assigned to both tiers.

We use a doubly-linked list for withdrawal requests.
When we remove a tier-two address, we iterate the list and delete its requests.
Each node in the list represents a request and includes: 
(1) the amount (a positive integer) we wish to draw from Phoenix's balance, 
(2) the recipient's network address, 
(3) the block number at which the request was made (to know when it can be withdrawn) and
(4) who initiated it (to allow only the second-tier address that initiated the request to cancel it and not other tier-two addresses).
 
Each node in the linked list has an ID. 
That ID is a cyclic counter, used to identify the request when we want to withdraw it or cancel it. 
As looking up a node in a linked list has an $O(n)$ time complexity, we use a mapping that gets an ID and returns a pointer to a node in the linked list.
Each node includes the ID of the previous and next nodes in the linked list. 
These IDs can then be looked up using the map. 
This data structure allows out-of-order insertion and deletion at $O(1)$ complexity. 
An illustration of the ledger's structure can be found in \Cref{App:Ledger}.

Removing all withdrawal requests of a deleted tier-two address still requires $O(n)$ operations as it is done by traversal over the entire linked list. 
If the number of nodes grows too big, traversing the list might be more expensive than the Ethereum gas limit~\cite{reynaldo_ethereum_2020}, rendering Phoenix unusable.
The solution is to monitor the request list's size and forbid it to grow over a maximal size limit. 
That limit is set at the contract's creation time.

The final problem is implementing the cancellation of all requests in the list at $O(1)$. 
Setting the head to point to null (ID 0) will make the old transactions inaccessible at list traversal. 
However, they could still be accessed from the map. 
The solution is to save an integer field, \lastid, which holds the largest ID of a request deleted by the latest cancel all requests operation. 
Transactions receive ID numbers in ascending order. 
Whenever we cancel all requests, we update \lastid\ to be the last given ID. 
Whenever we try to access a transaction, we return failure if its ID is lesser or equal to \lastid. 
There will be an eventual collision in id values when \lastid\  reaches an overflow. 
Nevertheless, that requires Phoenix to have $2^{256}$ withdrawal requests over its lifetime, an unrealistic scenario.

We tested the contract using Ganache~\cite{noauthor_trufflesuite/ganache-cli_2020} and Truffle~\cite{noauthor_trufflesuite/truffle_2020} with both black and white box JavaScript scripts. 
The code is released as open source.\footnote{\url{https://anonymous.4open.science/r/3b962760-6c42-4ecb-9325-d082d8b1b50b/}}


    \section{Formally Verifying the Correctness of the Code} 
    \label{sub:verification}

This section describes the process of formally verifying the code using the Certora
Prover~\cite{certora_homepage}. 
The tool takes as input (1) the EVM bytecode of the contract and (2) a specification describing the correctness properties of the contract.
It produces a report describing which correctness properties are valid and which are violated.
When a violated property is detected, the tool also produces a counterexample of an input scenario in which the property is violated.
The Certora Prover is based on well-established verification techniques in the line of Floyd-Hoare logic~\cite{hoare1969axiomatic}.
Other examples of similar tools include Dafny~\cite{leino2010dafny} and Why3~\cite{why3}.
In this technique, the correctness properties and the code are converted into a set of logical constraints. 
They are satisfied if there is an input leading to a property violation. 
These logical constraints can be solved using off-the-shelf SMT solvers such as Z3~\cite{z3} and CVC4~\cite{cvc4}.


        \subsection{Formal Specification of Phoenix} \label{sub:FormalSpec}
        
We begin by formalizing the specification presented in \Cref{sec:ModelAndSpec}.

    \bpar{Network and participants}
We assume the cryptocurrency's network is based on a blockchain, modeled as a series of \emph{blocks} $\period_1, \period_2, \dots$ .
The blocks are ordered chronologically.
A network address represents a private-public key pair.
A user who knows the private key is owning the address.
A participant can interact with Phoenix by using one of their addresses.
The interaction is done by taking one of the \emph{actions} defined below.
Actions might have limitations on which addresses can invoke them.
We assume a single block \period\ contains a single action.

Network addresses can be tier-one, tier-two, or neither.
The tier-one and tier-two address groups are disjoint.
Addresses not in any tier are called \emph{unprivileged} addresses.
We assume all participants can generate unprivileged addresses at will.
A participant could be honest or adversary.
An adversary might have network addresses with tier-one or tier-two privileges.
In particular, an adversary might own addresses also owned by other honest participants.

        \bpar{Phoenix}
A Phoenix contract, denoted~\phoenix, is a six-tuple $\phoenix = (\funds, \delay, \T1, \T2, \unlock, \allrequests)$.
\funds~is a non-negative integer representing the amount of cryptocurrency stored in~\phoenix. 
\delay~denotes delay, a positive quantity measured in network time. 
$\T1=\{\T1^1, \T1^2,\dots,\T1^m\}$ is a non-empty finite set of tier-one addresses. 
$\T2=\{\T2^1, \T2^2,\dots,\T2^n\}$ is a finite set of
tier-two addresses, which can be empty. 
\T1 and \T2 are disjoint.
\unlock~denotes a block, after which the contract is unlocked for cryptocurrency transfers.

\allrequests~is a finite set of withdrawal \emph{requests} $\request_1, \request_2,\dots,\request_l$.
A withdrawal \emph{request} $\request_i$ is a five-tuple $\request_i=(id_i, amount_i, recipient_i, creation_i, initiator_i)$.
$id_i$ is a unique identifier for the request.
$amount_i$ is a positive amount of coins that is to be withdrawn, where $\sum_{1 \leq i \leq |\allrequests|} amount_i \le  \funds$.
$recipient_i$ is a network address to which the money should be transferred. 
$creation_i$ is the block number at which $\request_i$ was added to \allrequests.
$initiator_i$ is a \T2 address.

        \bpar{Actions}
We list all actions in \Cref{Ta:Actions}.
Each action can be invoked by a participant only if they use a network address with the required privilege.
Note that any participant who possesses a \T1 address can invoke \T2 privileged actions by generating an unprivileged address and adding it as a new \T2 address.

\begin{table}[!t]
\renewcommand{\arraystretch}{1.2}
\centering
\begin{tabular}[t]{l>{\raggedright}p{0.11\linewidth}>{\raggedright\arraybackslash}p{0.6\linewidth}}
\toprule
Action & Address & Description\\
\midrule
Deposit & Any & Increases the value of \funds.\\

Request & \T2 & Adds a new request $\request_k$ to \allrequests. Only \T2 users can take this action. If invoked by address $\T2^j$, then $initiator_k = \T2^j$. $creation_k$ is equal to the current block number as tracked by the network. This action fails unless $\sum_{1 \leq i \leq |\allrequests|} amount_i + amount_k \le  \funds$.\\

Withdraw & Any & Removes a request $\request_k$ from \allrequests. Decreases \funds\ by $amount_k$. This action can be invoked only if the current block's number is greater than $creation_k + \delay$ and greater than \unlock.\\

Cancel request & \T1 & Removes a request $\request_k$ from \allrequests.\\

Cancel all requests & \T1 & After this action is invoked, $\allrequests = \phi$.\\

Cancel self request & \T2 & Removes a request $\request_k$ from \allrequests, if $initiator_k$ is the same address as the \T2 that invoked this action.\\

Lock & \T1 & Increases \unlock.\\

Add a \T1 address & \T1 & Adds a user address $\T1^k$ to \T1 if $\T1^k \notin \T1 \cup \T2$.\\

Add a \T2 address & \T1 & Adds a user address $\T2^k$ to \T2 if $\T2^k \notin \T1 \cup \T2$.\\


Remove a \T2 address & \T1 & Removes a $\T2^k$ address from \T2 and all $\request_i \in \allrequests$ where $initiator_i = \T2^k$.\\

\bottomrule
\end{tabular}
\caption{\label{Ta:Actions}%
Participant actions on Phoenix}
\end{table}%

\subsection{Verification}

    \bpar{Correctness Properties} \label{Sub:CorrectnessProperties}
The formal verification tool takes as an input a mathematically precise specification. 
Our specification constitutes of a list of \emph{safety properties}~\cite{liveness_safety}. 
A verified safety property indicates that a "bad" state is unreachable.
We will describe the properties below in English and provide a formalization in \Cref{app:informal_prop}. 
The specification files we used can be found at the code's repository in the link in \Cref{Se:Architrcture}.

The properties were divided into different layers, depending on the implications of their violation.
The \emph{base layer} properties are: (1.1) requests cannot be withdrawn unless they have been kept in Phoenix's records for an amount of time equal or greater than Phoenix's delay; (1.2) a tier-one address can cancel any request at any time; (1.3) Phoenix's delay can never change; (1.4) There is no way to remove a tier-one address and (1.5) Phoenix cannot be destroyed unless it is empty.
Property 1.5 is necessary because in Ethereum~\cite{solidity_contracts_documentation} when a contract is destroyed, the sender of the destroy transaction can send all the funds in the contract to any address they choose.

The \emph{key separation layer} properties are: (2.1) an address cannot have both tier-one and tier-two privileges; (2.2) tier-one addresses can only be added by a tier-one address; (2.3) only a tier-two address can spend money; and (2.4) a tier-two address can only remove self-initiated requests.
These properties ensure the separation between the two user tiers.

The \emph{recovery layer} properties are: (3.1) money cannot leave Phoenix while it is locked; (3.2) the unlock time can only be postponed; (3.3) only tier-one addresses can lock Phoenix; (3.4) only tier-one addresses can remove tier-two addresses; and (3.5) only tier-one addresses can add tier-two addresses.
Recovery layer properties ensure the correctness of Phoenix's recovery mechanisms as detailed in \Cref{sec:ModelAndSpec}.

The \emph{tier-one minimization layer} properties are: (4.1) there are always enough funds in Phoenix to pay for all requests; (4.2) Phoenix cannot send money to itself; (4.3) Phoenix cannot send money to the zero address; and (4.4) removing a tier-two address from Phoenix also removes all requests initiated by that address.
The tier-one minimization layer properties reduce the number of transactions that require a \T1 address, reducing its exposure.

Property 4.1 prevents a \emph{delay-evasion} attack as described in \Cref{sub:attacks}.
A Phoenix sending money to itself is a pure waste of resources.
In Solidity, the zero address does not belong to any entity.
Sending money to that address is akin to burning it~\cite{solidity_contracts_documentation}.
Properties 4.2 and 4.3 prevent erroneous transactions to those addresses, which would require \T1 intervention to cancel.
Regarding property 4.4, if removing a \T2 address would not remove all requests initiated by it, the \T1 address could remove them manually, one at a time.
However, manual deletion increases the probability of an error.

The vault contract presented by MES~\cite{moser2016bitcoin} only satisfies the properties in the Base Layer.
BLZ vaults~\cite{bartoletti2020bitcoin} also satisfy the properties in the key separation layer.
Phoenix satisfies all properties listed above.

\subsection{Attacks, Their Mitigation and Formal Verification}
\label{sub:attacks}

The automatic prover uncovered several bugs in our initial implementation. 
The most severe issue was a violation of property 4.1 (\Cref{Sub:CorrectnessProperties}), which we now describe. 

The bug is related to what we call a \emph{delay-evasion attack}. 
In this attack, the attacker adds one or more \emph{delay-evading} requests that cannot be completed due to insufficient funds at attacking time. 
Later, once sufficient funds are deposited, an immediate claim for the delay-evading requests could be made without waiting for the delay period.

Formally, let $\funds_j$ be the value of $\funds$ at time $\tau_j$.
At time $\tau_0$, a request $\request_k$ is made for an amount $k>\funds_0$.
If at time $\tau_1 > \tau_0 + \delay$, $\funds_1 \geq k$, the attacker claims $\request_k$ without waiting for a \delay\ amount of time after the money was deposited.
The attack can be executed with a set of requests $\request_{k_1}, \request_{k_2} \dots \request_{k_m}$ such that $\sum_{1 \leq i \leq m} amount_{k_i} > \funds_0$.

To prevent delay-evasion, we check for every new request if the sum of the amounts of existing requests and the new request is smaller than \funds.
If the condition does not hold, the request is aborted.
However, our code did not check if the sum of all requests overflows. 
EVM uses 256-bit wide integers~\cite{ethereum_yellow_paper} and we denote their maximal value as $\textit{MAX\_INT}$.

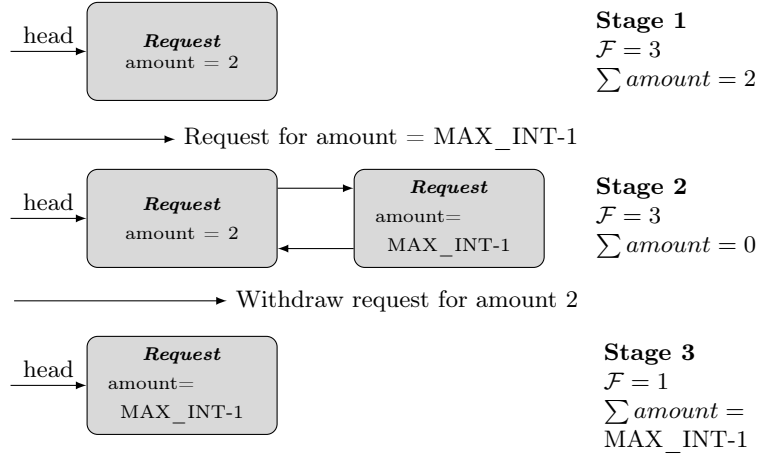
\begin{figure}[!t]
\centering
\tikzstyle{block} = [rectangle, draw, text width=6em, text centered, rounded corners, minimum height=4em, fill=gray!30, minimum width=2.5cm, align=center]
\tikzstyle{line} = [draw, -latex]

\begin{tikzpicture}[node distance=1cm, auto]
    \node (init) {};
    
    \node [block] (s1_head) {\scriptsize \textbf{\textit{Request}}\\
    amount = 2\\};
    \coordinate [left=1cm of s1_head] (UL);
    \path [line]   (UL) -- node [text width=2.5cm,midway,above,align=center ] {head} (s1_head);
    
    \node[align=left] (s1_text) at (6.5,0) {\textbf{Stage 1} \\ $\funds = 3$ \\ $\sum amount = 2$};
    
    \node[align=left] (s1_transition) [below left = 0.25 cm and 0 cm of s1_text] {Request for amount = MAX\_INT-1};
    
    \coordinate [left=2.15cm of s1_transition] (UT);
    \path [line]   (UT) -- node [text width=2.5cm,midway,above,align=center ] {} (s1_transition);
    
    \node [block] (s2_head) [below = 0.9cm of s1_head]{\scriptsize \textbf{\textit{Request}}\\
    amount = 2};
    \coordinate [left=1cm of s2_head.west] (ML);
    \path [line]   (ML) -- node [text width=2.5cm,midway,above,align=center ] {head} (s2_head);
    
    \node [block] (s2_overflow) [right=1cm of s2_head.east] {\scriptsize \textbf{\textit{Request}}\\
    amount= \newline MAX\_INT-1};
    
    \path [line] ([shift={(0,4mm)}]s2_head.east) -- node [text width=1cm, midway, above, align=center ] {} ([shift={(0,4mm)}]s2_overflow.west);
    \path [line] ([shift={(0,-4mm)}]s2_overflow.west) -- node [text width=1cm, midway, above, align=center ] {} ([shift={(0,-4mm)}]s2_head.east);
    
    \node[align=left] (s2_text) [below=0.9cm of s1_text] {\textbf{Stage 2} \\ $\funds = 3$ \\ $\sum amount = 0$};
    
    \node[align=left] (s2_transition) [below left = 0.2 cm and 0 cm of s2_text] {Withdraw request for amount 2};
    
    \coordinate [left=2.8cm of s2_transition] (MT);
    \path [line]   (MT) -- node [text width=2.5cm,midway,above,align=center ] {} (s2_transition);
    
    \node [block] (s3_head) [below = 0.9cm of s2_head]{\scriptsize \textbf{\textit{Request}}\\
    amount= \newline MAX\_INT-1};
    \coordinate [left=1cm of s3_head.west] (LL);
    \path [line]   (LL) -- node [text width=2.5cm,midway,above,align=center ] {head} (s3_head);
    
    \node[align=left] (s3_text) [below=0.9cm of s2_text] {\textbf{Stage 3} \\ $\funds = 1$ \\ $\sum amount = $ \\ MAX\_INT-1};
    
\end{tikzpicture}
\caption{\textit{Denial of Service attack. During stages 1 and 2, property 4.1 of \Cref{Sub:CorrectnessProperties} holds. At stage 3, the property is violated. If, after stage 3, a user tries to request the transfer of 1 coin, the action will fail despite Phoenix having sufficient funds.}} \label{fig:DoS}
\end{figure}

This weakness can be exploited in two ways: a delay-evasion attack and a denial of service (DoS) attack.
We start by describing the DoS.
We assume a non-empty~\allrequests\ and denote $K = \sum_{1 \leq i \leq |\allrequests|}amount_i > 0$, where $K\le \funds$.
An attacker requests to withdraw $\textit{MAX\_INT}-K+1$.
An overflow occurs and the request is accepted, as now $\sum_{1 \leq i \leq |\allrequests|}amount_i + \textit{MAX\_INT}-K+1=0$.
Then, the attacker waits until a request for an amount of $L\le K$ is claimed.
The new sum over \allrequests\ is $\textit{MAX\_INT}-L+1$.
Hence, any request with an amount smaller than $L$ will not overflow.
If the sum is greater than \funds\ (which is the realistic case), such requests must fail.
An example scenario where $K=2,\ L=2$ is illustrated in \Cref{fig:DoS}.

\begin{figure}[!t]
\centering
\tikzstyle{block} = [rectangle, draw, text width=5.4em, text centered, rounded corners, minimum height=4em, fill=gray!30, minimum width=1cm, align=center]
\tikzstyle{line} = [draw, -latex]

\begin{tikzpicture}[node distance=1cm, auto]
    \node (init) {};
    
    \node [block] (s1_head) [text width=5em] {\scriptsize \textbf{\textit{Request}}\\
    amount = 2\\};
    \coordinate [left=1cm of s1_head] (UL);
    \path [line]   (UL) -- node [text width=2cm,midway,above,align=center ] {head} (s1_head);
    
    \node[align=left] (s1_text) at (7.5,0) {\textbf{Stage 1} \\ $\funds = 2$ \\ $\sum amount = 2$};
    
    \node[align=left] (s1_transition) [below left = 0.3 cm and 0 cm of s1_text] {Request for amount = MAX\_INT-1};
    
    \coordinate [left=2.9cm of s1_transition] (UT);
    \path [line]   (UT) -- node [text width=2.5cm,midway,above,align=center ] {} (s1_transition);
    
    \node [block] (s2_head) [below = 1.1cm of s1_head, text width=5em]{\scriptsize \textbf{\textit{Request}}\\
    amount = 2};
    \coordinate [left=1cm of s2_head.west] (ML);
    \path [line]   (ML) -- node [text width=2.5cm,midway,above,align=center ] {head} (s2_head);
    
    \node [block] (s2_overflow) [right=0.5cm of s2_head.east] {\scriptsize \textbf{\textit{Request}}\\
    amount= \newline MAX\_INT-1};
    
    \path [line] ([shift={(0,4mm)}]s2_head.east) -- node [text width=1cm, midway, above, align=center ] {} ([shift={(0,4mm)}]s2_overflow.west);
    \path [line] ([shift={(0,-4mm)}]s2_overflow.west) -- node [text width=1cm, midway, above, align=center ] {} ([shift={(0,-4mm)}]s2_head.east);
    
    \node[align=left] (s2_text) [below=1.1cm of s1_text] {\textbf{Stage 2} \\ $\funds = 2$ \\ $\sum amount = 0$};
    
    \node[align=left] (s2_transition) [below left = 0.3 cm and 0 cm of s2_text] {Request for amount = 2};
    
    \coordinate [left=4.5cm of s2_transition] (MT);
    \path [line]   (MT) -- node [text width=2.5cm,midway,above,align=center ] {} (s2_transition);
    
    \node [block] (s3_head) [below = 1.1cm of s2_head, text width=5em]{\scriptsize \textbf{\textit{Request}}\\
    amount = 2};
    \coordinate [left=1cm of s3_head.west] (LL);
    \path [line]   (LL) -- node [text width=2cm,midway,above,align=center ] {head} (s3_head);
    
    \node [block] (s3_overflow) [right=0.5cm of s3_head.east] {\scriptsize \textbf{\textit{Request}}\\
    amount= \newline MAX\_INT-1};
    
    \path [line] ([shift={(0,4mm)}]s3_head.east) -- node [text width=1cm, midway, above, align=center ] {} ([shift={(0,4mm)}]s3_overflow.west);
    \path [line] ([shift={(0,-4mm)}]s3_overflow.west) -- node [text width=1cm, midway, above, align=center ] {} ([shift={(0,-4mm)}]s3_head.east);
    
    \node [block] (s3_ambush) [right=0.5cm of s3_overflow.east, text width=5em] {\scriptsize \textbf{\textit{Request}}\\
    amount = 2};
    
    \path [line] ([shift={(0,4mm)}]s3_overflow.east) -- node [text width=1cm, midway, above, align=center ] {} ([shift={(0,4mm)}]s3_ambush.west);
    \path [line] ([shift={(0,-4mm)}]s3_ambush.west) -- node [text width=1cm, midway, above, align=center ] {} ([shift={(0,-4mm)}]s3_overflow.east);
    
    \node[align=left] (s3_text) [below=1.1cm of s2_text] {\textbf{Stage 3} \\ $\funds = 2$ \\ $\sum amount = 2$};
    
    \node[align=left] (loop_icon) [below left = 0.6cm and 0.8 cm of s2_text.east]{\includegraphics[width=1cm]{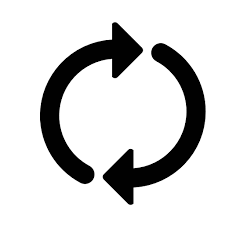}};
    
\end{tikzpicture}
\caption{\textit{Delay-evasion attack. The actions in stages 2 and 3 preserve the sum of the amounts but can be used to withdraw freshly deposited funds. For example, by repeating stages 2 and 3 $n$ times, a deposit of up to $2n$ can be withdrawn by the attacker immediately.}}
\label{fig:Ambush}
\end{figure}
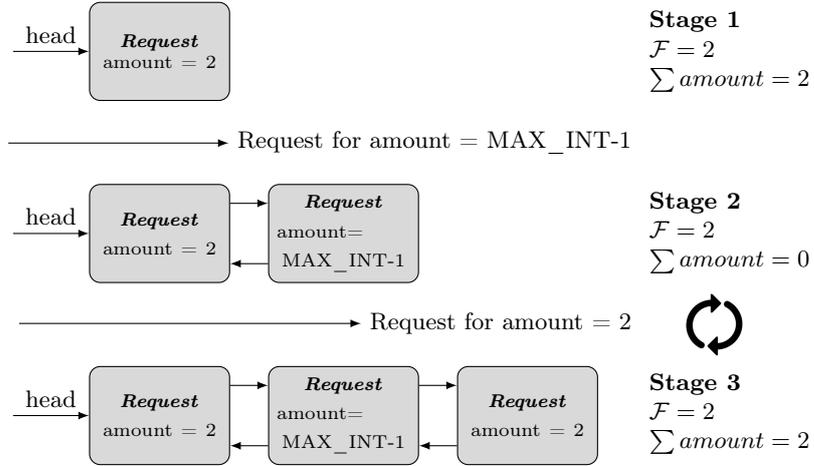

The bug's second manifestation is in a \emph{delay-evasion} attack.
We again assume a non-empty ledger, denote $K = \sum_{1 \leq i \leq |\allrequests|}amount_i > 0$ where $K \le \funds$.
The attack begins with a request to withdraw $\textit{MAX\_INT}-K+1$.
Overflow occurs, and the request is accepted as $\sum_{1 \leq i \leq |\allrequests|}amount_i + \textit{MAX\_INT}-K+1 = 0$.
Then, the attacker issues two consecutive requests: first for an amount equal to $L \le \funds$, and second to an amount equal to $\textit{MAX\_INT}-L+1$.
After these two requests, the sum over \allrequests\ is still equal to 0.
The attacker can continue to issue such pairs of requests, with different values for $L$, preserving the sum at value 0.
The first requests in those pairs are the delay-evading requests.
Suppose at any point in the future, after \delay\ blocks since the attacker's actions elapsed, a value of $L$ or greater is deposited. 
In that case, the delay-evading request for $L$ can be withdrawn immediately after the deposit.
An example scenario where $K=2, \funds = 2, L=2$ is illustrated in \Cref{fig:Ambush}.

The bug fix is to check explicitly if an overflow occurs in computing the sum of amounts when adding a new request.
If an overflow is detected, we reject the request.
The formal verification tool proved the correctness of the code after the fix.

\subsection{Limitations of the Verification}

We have only proved that the properties are correct for cases where the number of loop iterations is one or less.
The increase from one iteration to two introduces many more counterexamples, rising from the need to verify invariants of a linked list (see \Cref{Se:Architrcture}).
For example, we need to verify that the list does not contain cycles or branches, and that forward and backward arrows are consistent. 

\section{Conclusion} \label{sec:conclusion}
We presented Phoenix, a wallet contract that allows for full recovery after the most common forms of attack and partial recovery against the rest.
We defined a formal model for Phoenix, implemented it in Solidity, and formally verified it.

The mechanisms presented in Phoenix can be extended and applied to additional use-cases. 
One is to allow for rate-limited withdrawals without delays~-- permit small amounts to be withdrawn immediately up to a limit, with that limit reset in regular intervals. 

Phoenix's hierarchical structure can be utilized in organizations, enabling them to assign multiple employees with withdrawal privileges that are revoked if they leave the organization. 
Phoenix can be further extended to more hierarchy levels~-- a dynamic version of HD-wallets~\cite{hdwallets}.

An additional use-case is a digital will~\cite{seres2020cryptowills,rice2019deceased}. 
The owner initiates a Phoenix contract keeping a \T1 address for herself and adding her heirs as \T2 addresses.
As long as the owner is alive, the heirs cannot withdraw money from Phoenix; she can cancel such requests.
Once the owner is deceased, the heirs can withdraw the money freely.
This can be further extended to support dynamic allocation to multiple heirs. 

\newpage
\bibliography{citations} 
\bibliographystyle{plainurl}

        \pagebreak
        \appendix

        \section{Comparison Between User Tiers}
\label{App:Tiers}

We summarize the roles of the two tier addresses in \Cref{AppTa:Tiers}.

\begin{table}[!ht]
\renewcommand{\arraystretch}{1.25}
\centering
\begin{tabular}[t]{l>{\raggedright}p{0.4\linewidth}>{\raggedright\arraybackslash}p{0.3\linewidth}}
\toprule
& Tier-One Address & Tier-Two Address\\
\midrule
\textbf{Main function} & cancel unauthorized transfers & money transfer \\
\textbf{Usage frequency} & rarely & daily \\ 
\textbf{Appointed by} & a tier-one address & a tier-one address \\
\textbf{Degradation} & impossible & by a tier-one address \\
\textbf{Request cancellation} & can cancel any request & can only cancel requests they initiated \\
\textbf{Lock} & can lock Phoenix & cannot lock Phoenix \\
\textbf{Wallet termination} & can only terminate an empty Phoenix & cannot terminate \\

\bottomrule
\end{tabular}
\caption{Comparison of user tiers}
\label{AppTa:Tiers}
\end{table}%
\clearpage

        \section{Comparison of Monetary Losses Between Phoenix and a Vault Contract}

\label{App:Venn}

\begin{figure}[!ht]
\centering
\begin{tikzpicture}
  \tikzset{venn circle/.style={draw,circle,minimum width=2.5cm}}


  \node [venn circle, label=below:{Tier 1}] (tier1_loss_vault) at (0,0) {$\epsilon$};
  \node [venn circle, label=below:{Tier 2}] (tier2_loss_vault) at (0:1.6cm) {\funds};
  \node (all_keys_lost_vault) at (barycentric cs:tier1_loss_vault=1/2,tier2_loss_vault=1/2) {\funds};
  
  \node[] (key_loss_vault) [below=1.6cm of all_keys_lost_vault]{\large{Key Loss}};

  \node [venn circle, label=below:{Tier 1}] (tier1_theft_vault) at (5,0) {\funds};
  \node [venn circle, label=below:{Tier 2}] (tier2_theft_vault) at (0:6.6cm) {\funds*};
  \node (all_keys_stolen_vault) at (barycentric cs:tier1_theft_vault=1/2,tier2_theft_vault=1/2) {\funds};
  
  \node[] (key_theft_vault) [below=1.6cm of all_keys_stolen_vault]{\large{Key Theft}};
  
  \node[] (BLZ)[below right=0.1cm and 0.0cm of key_loss_vault]{\large{\textbf{BLZ Vault}}};


  \node [venn circle, label=below:{Tier 1}] (tier1_loss_phoenix) at (0,-5) {$\epsilon$};
  \node [venn circle, label=below:{Tier 2}] (tier2_loss_phoenix) at (1.6cm,-5cm) {$\mathlarger{\mathlarger{\mathlarger{\bm{\epsilon}}}}$};
  \node (all_keys_lost_phoenix) at (barycentric cs:tier1_loss_phoenix=1/2,tier2_loss_phoenix=1/2) {\funds};
  
    \node[] (key_loss_phoenix) [below=1.6cm of all_keys_lost_phoenix]{\large{Key Loss}};

  \node [venn circle, label=below:{Tier 1}] (tier1_theft_phoenix) at (5,-5) {\funds};
  \node [venn circle, label=below:{Tier 2}] (tier2_theft_phoenix) at (6.6,-5cm) {$\mathlarger{\mathlarger{\mathlarger{\bm{\epsilon}}}}$};
  \node (all_keys_stolen_phoenix) at (barycentric cs:tier1_theft_phoenix=1/2,tier2_theft_phoenix=1/2) {\funds};
  
  \node[] (key_theft_phoenix) [below=1.6cm of all_keys_stolen_phoenix]{\large{Key Theft}};
  
  \node[] (Phoenix)[below right=0.1cm and 0.0cm of key_loss_phoenix]{\large{\textbf{Phoenix Vault}}};
\end{tikzpicture}

\caption{\textit{Monetary loss by the owner in various cases where all keys of a given tier are lost in Phoenix and a BLZ vault~\cite{bartoletti2020bitcoin}. In a BLZ vault contract, when a tier-two key is stolen, a wealthy attacker that does not care about her losses could cause the depletion of the entire contents of the vault via attrition. Scenarios in which the Phoenix yields better results than BLZ vaults are in bold.}} \label{AppFig:Venn}

\end{figure}
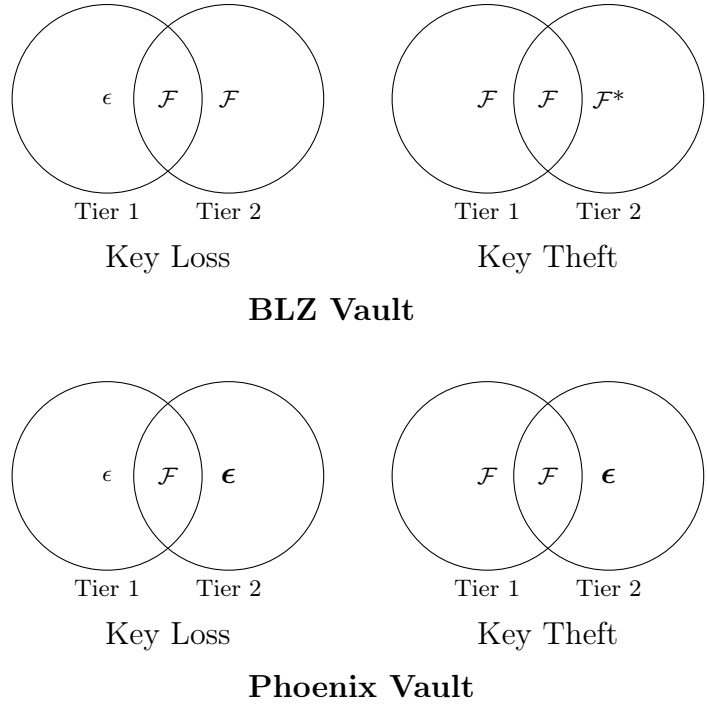

We compare the monetary losses to the owner due to attacks and benign key losses between Phoenix and a BLZ vault contract~\cite{bartoletti2020bitcoin} in \Cref{AppFig:Venn}.
When comparing attacks, we assume an asymmetric situation where the attacker is malicious and does care for her self-incurred monetary losses. 

\clearpage

        \section{Ledger Structure}

\label{App:Ledger}

\begin{figure}[!h]
\centering
\includegraphics[width=\textwidth]{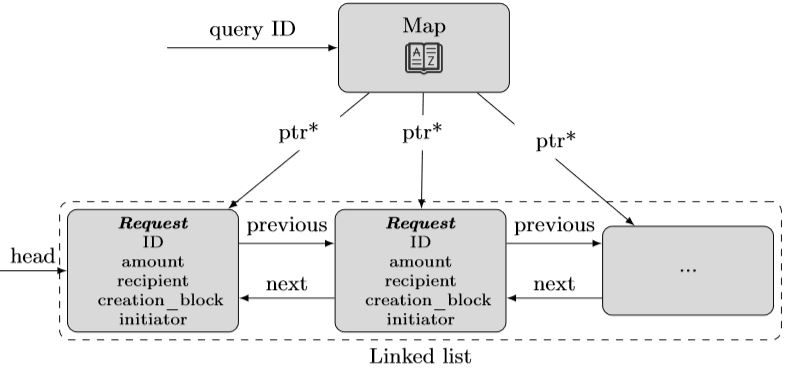}
\caption{Ledger structure} \label{Fig:LedgerStruct}
\end{figure}

We present the architecture of Phoenix's Ledger library in \Cref{Fig:LedgerStruct}.

\clearpage


        \section{Properties of Phoenix}
\label{app:informal_prop}

A summary of the properties in English appears in \Cref{Ta:informal_prop}.
Their formal definitions appear in \Cref{Ta:formal_prop}.


\begin{table}[ht]
\centering
\begin{tabular}[t]{l>{\raggedright\arraybackslash}p{0.95\linewidth}}

\toprule
\multicolumn{2}{c}{Base layer} \\
\midrule

1.1 & Requests cannot be withdrawn unless they have been kept in Phoenix's records for an amount of time equal or greater than Phoenix's delay. \\

1.2 & A tier-one address can cancel any request at any time. \\

1.3 & Phoenix's delay can never change. \\

1.4 & There is no way to remove a tier-one address. \\

1.5 & Phoenix cannot be destroyed unless it is empty. \\

\midrule
\multicolumn{2}{c}{Key separation layer} \\
\midrule

2.1 & An address cannot have both tier-one and tier-two privileges. \\

2.2 & Tier-one addresses can only be added by a tier-one address. \\

2.3 & Only a tier-two address can spend money. \\

2.4 & A tier-two address can only remove self-initiated requests. \\

\midrule
\multicolumn{2}{c}{Recovery layer} \\
\midrule

3.1 & Money cannot leave Phoenix while it is locked. \\
3.2 & The unlock time can only be postponed. \\
3.3 & Only tier-one addresses can lock Phoenix.\\
3.4 & Only tier-one addresses can remove tier-two addresses.\\
3.5 & Only tier-one addresses can add tier-two addresses.\\

\midrule
\multicolumn{2}{c}{Tier-one minimization layer} \\
\midrule
4.1 & There are always enough funds in Phoenix to pay for all requests.\\
4.2 & Phoenix cannot send money to itself.\\ 
4.3 & Phoenix cannot send money to the zero address.\\ 
4.4 & Removing a tier-two address from Phoenix also removes all requests initiated by that address.\\

\bottomrule
\end{tabular}
\caption{Phoenix correctness properties}
\label{Ta:informal_prop}
\end{table}%
\clearpage

\begin{table}[ht]
\centering
\begin{tabular}[t]{l>{\raggedright\arraybackslash}p{0.95\linewidth}}

\toprule
\multicolumn{2}{c}{Base layer} \\
\midrule

1.1 & When the current block number is smaller than $creation_k + \delay$, withdrawal of $r_k$ will fail. \\

1.2 & Any withdrawal request $\request \in \allrequests$ can at any time be cancelled by any \T1 address. \\

1.3 & If at block number $\tau_0$, $\delay_0=\delay$, then at block number $\tau_1$, $\delay_1 = \delay$. \\

1.4 & If at block number $\tau_0$, $\T1^k \in \T1$, then $\forall \tau_1 > \tau_0,\ \T1^k \in \T1$. \\

1.5 & \phoenix\ can only be destroyed if $\funds=0$. \\

\midrule
\multicolumn{2}{c}{Key separation layer} \\
\midrule

2.1 & $\T1 \cap \T2 = \phi$. \\

2.2 & Only \T1 addresses can add \T1 addresses. \\

2.3 & Only \T2 addresses can take the \emph{request} action. \\

2.4 & A $\T2^j \in \T2$ address cannot remove a request $\request_k \in \allrequests$ unless $inititator_k=\T2^j$. \\

\midrule
\multicolumn{2}{c}{Recovery layer} \\
\midrule

3.1 & If at block number $\tau_0 < \unlock$ Phoenix contains $\funds = \funds_0$, then  $\forall \tau_0 \leq \tau_1 < \unlock$, $\funds_1 \geq \funds_0$. \\
3.2 & If at block number $\tau_0$ the unlock block number is $\unlock = \unlock_0$, then $\forall \tau_1 \geq \tau_0,\ \unlock_1 \geq \tau_0$. \\
3.3 & Only a \T1 address can take the \emph{lock} action. \\
3.4 & Only \T1 addresses can delete \T2 addresses. \\
3.5 & Only \T1 addresses can add \T2 addresses. \\

\midrule
\multicolumn{2}{c}{Tier-one minimization layer} \\
\midrule

4.1 & $\sum_{1 \leq i \leq |\allrequests|} amount_i \le  \funds$. \\

4.2 & $\forall \request_i \in \allrequests,\ recipient_i \neq \phoenix$. \\
4.3 & $\forall \request_i \in \allrequests,\ recipient_i \neq address(0)$. \\ 
4.4 & $\forall \request_i \in \allrequests,\ initiator_i \in \T2$. \\

\bottomrule
\end{tabular}
\caption{Phoenix formal correctness properties}
\label{Ta:formal_prop}
\end{table}%

\end{document}